# Inertial System and Special Relativity

# Finite Geometrical Field Theory of Matter Motion Part One

Xiao Jianhua

Natural Science Foundation Research Group, Shanghai Jiaotong University

Shanghai, P.R.C

**Abstract:** Special relativity theory is well established and confirmed by experiments. This research establishes an operational measurement way to express the great theory in a geometrical form. This may be valuable for understanding the underlying concepts of relativity theory. In four-dimensional spacetime continuum, the displacement field of matter motion is measurable quantities. Based on these measurements, a finite geometrical field can be established. On this sense, the matter motion in physics is viewed as the deformation of spacetime continuum. Suppose the spacetime continuum is isotropic, based on the least action principle, the general motion equations can be established. In this part, Newton motion and special relativity are discussed. Based on the finite geometrical field theory of matter motion, the Newton motion equation and the special relativity can be derived simply based on the isotropy of spacetime continuum and the definition of inertial system. This research shows that the Lorentz transformation is required by both of the inertial system definition and the time gauge invariance for inertial systems. Hence, the special relativity is the logic conclusion of time invariance in inertial system. The source independent of light velocity supports the isotropy of inertial system rather than the concept of proper time, which not only causes many paradox, such as the twin-paradox, but also causes many misunderstanding and controversial arguments. The singularity of Lorentz transformation is removed in other parts of finite geometrical field theory, where the gravity field, electromagnetic field, and quantum field will be discussed with the time displacement field.

## 1. Introduction

Newton mechanics and special relativity theory is well established and confirmed by experiments. What is their intrinsic relation? There are huge papers on the special relativity theory and its relationship with Newton mechanics. Howerver, the ruler and clock in Einstein's theory cause many contraversial arguments about the time concept and the phisical intrinsic maening of Lorentz transformation. The aparant shortage of Minkowski spacetime is the negativity of time gauge (or space gauge). However, the Minkowski spacetime has been accepted as the bases for many researches. This is a fact. But this fact cannot be used to prove that the Minkowski spacetime is the unique solution for Lorentz transformation. This paper will show that there is another way to introduce the Lorentz transformation in a natural way following the traditional way of inertial system. Although the treatment expressed in this paper is very traditional in the operational sense of measurements, the concept of general relativity is fully accepted.

In this research, the space gauge and time gauge are viewd as the basic feature of matter motion. The Riemann geometry in four-dimenssional spacetime is adapted as the basic geometry of matter motion. There are many reasons for this view-point. It is believed that for typical matter motion there is no need to introduce too complicated geometrical structures, as the basic problems are well solved by simple geometry in each branch of physics.

This research will not review the related researches as there are many excelent review articles available. The puppse of this research is to establish the intrinsic form of matter motion by operational measurement geometrical method. The physical laws are derived from the least action principle proposed by Einstein and the inertial definition established by Newton. Of cource, some ad hoc principles must be adapted. The research exposes the intrinsic relation among basic physical laws rather than inventing some laws. The research focus on reformulating the basic physical laws in an operational way which, the author believes, will make it be easy for one to understand the complecated physical world in tradition-like way.

For matter motion, the basic operational measurements are the geometrical variation of spacetime continuum. Hence, an intrinsic geometrical structure should be established firstly (this believe is contradict with the concepts of looking the underlying geometrical structures for known physical motion laws). Secondly, the underlying physical laws are established to describe the cause of geometrical variation and their intrinsic relationship. This forms the basic believe of the whole research, which will be expalined by three basic papers of the research.

In this research, the first principle is the isotropy of spacetime (the isotropy of inertial mass is implied by this principle). The second principle is the least action principle (the Einstein field equation is implied by this principle). As the research will show that there are no contradicts or theoretic gap among classical mechanics, general reletivity (gravitation), and the quantum mechanics. The research will be roughly devided into three parts. (1). The classical Newton mechanics is the aproximation when the time is constant flow. When the time is non-constant flow, the Lorentz transformation must be introduced to meet the requirement of inertial system. This established the intrinsic relationship between the Newton mechanics and special relativity. The gravity field is the intrinsic time gauge variation in space, that has been well explained in general relativity by the Schwarzshild solution. For such a kind of matter motion, the geometric feature of matter motion is commutative. (2). The pure time flow variation field is the electromagnetic field, which will be fully discussed in another paper. For such a kind of matter motion, the non-commutative geometry will play an important role. This topic will be discussed in a special paper (part two) later. (3). When the time gauge variation is coupled with the space gauge variation, quantum mechanics will be established. Hence, the quatum mechanics is the most complecated matter motion. However, they are intrinsicly related with the simple matter motion in a unified field theory form. This topic will be discussed in part three paper of this research.

In this paper, the main topic is the intrinsic relation between the inertial system and special relativity. The research shows that the isotropy of inertial mass is implied by the principle of spacetime isotropy. This will be explained in the main body of this paper.

The definition of inertial system reqires that the physical laws be identical for inertial system. The research will show that for two relative moving inertial system this reqirement will demand the Lorentz transformation. The source independent of light speed supports the the physical laws be identical for inertial system. So, this research shows the Lorentz transformation as the result of inertial system. Therefore, the Lorentz transformation is not basic and cannot be taken as the base of physics. That is to say the Minkowski spacetime is not a good choice for physics, although it is correct for the inertial motion and the commutative geometric theory of gravitation field and quantum field.

Based on this general view point, this paper firstly introduces the motion transformation. For its relation with other geometrical theory the author has given out a simple explaination [1]. Secondly, the inertia definition is fully examined to show the Lorentz transformation is reqiured by the definition of inertia. Then the least action principle is used to establish general motion equations for matter motion. The paper shows that the simplest motion of matter is the Newton motion equations. Finally, based on the principle of physical laws be identical for inertial system, the gavitation field is established. The result shows that inertial mass is identical with the gravitational mass only when the quntum effects and elctromagnetic field can be ignored. Therefore, the identicy of inertial mass and gravitational mass can not be taken as basic principle for unified field theory.

**2. Matter Motion Transformation**

Although the absolute space and time (world time) defined by Newton is not accepted in physics, but it is a good concept for discuss matter motion in the inertial system. In this research, the absolut spacetime is defined as vacum or ether. The absolute spacetime has the ruler and clock as phisical gauge. The research will used the absolute space and time to parameterizing matter in its absolute motionless state.

Introducing "world time" as time coordinator parameter in the Riemannian geometry attached with matter (be defined as four-dimensional co-moving coordinator system). By this way, a matter point (event) is identified geometrically by its intrinsic four-dimensional position. Its motion will determine its actual gauge field which is measured by "oberservor" in conventional geometry. Hence, the matter motion will be expressed by the variation of actual gauge field. As the variation of gauge field can be expressed by the transformation of basic vectors of four-dimensional co-moving coordinator system, the motion transformation is defined to describe the matter motion.

The Riemannian geometry attached with matter in three-dimensional co-moving coordinator system has been established by Chen Zhida [2-3]. His research shows that the Riemannian geometry attached with matter is determined by the displacement field of matter, which is measured by "observor" in conventional geometry.

This paper defines the spacetime geometry attached with "observer" as "vacuum" geometry and the spacetime geometry attached with matter as the Riemannian geometry. The initial spacetime geometry attached with matter is taken as refference geometry (named as initial co-moving coordinator system), and the current spacetime geometry attached with matter (named as current co-moving coordinator system) is determined by the four-dimensional displacement field measured in initial spacetime geometry.

In the initial co-moving coordinator system, matter is identified by four-dimensional co-moving coordinators. For the matter, the four-dimensional co-moving coordinators are invariant while the actual gauge field is varying with matter motion (hence the four-dimensional co-moving coordinator system is varying with matter motion).

The initial four-dimensional co-moving coordinator system is defined by anti-covariant coordinators ($x^1, x^2, x^3, x^4$) and initial basic vectors ($\vec{g}_1^0, \vec{g}_2^0, \vec{g}_3^0, \vec{g}_4^0$). Where, $x^4$ is taken as time coordinator. From physical consideration, the initial co-moving coordinator system can be defined as the standard three-dimensional space adding time-dimension, in which following conditions are met:

$$\vec{g}_i^0 \cdot \vec{g}_j^0 = \begin{cases} 0, i \neq j \\ 1, i = j \neq 4 \\ c^2, i = j = 4 \end{cases} \quad (1)$$

where, $c$ is light speed in "vacuum". The difference with the Riemannian geometry of Einstein's theory is that time gauge in here is real rather than imaginary.

By such a sellection, matter is defined in standard physical measuring system at initial time of motion.

In physics, matter motion can expressed by displacement field $u^i$ measured in standard physical measuring system. By above sellection, the displacement field $u^i$ is also defined in the initial four-dimensional co-moving coordinator system.

For any measuring spacetime, the matter in discussing has invariant coordinators, but its local basic vectors are changed into ($\vec{g}_1, \vec{g}_2, \vec{g}_3, \vec{g}_4$). According to Chen Zhida's research [2-3], for incremental motion of matter, there are relationship between initial basic vectors and current basic vectors:

$$\vec{g}_i = F_i^j \vec{g}_j^0 \quad (2)$$

and the transformation tensor $F_j^i$ is determined by equation:

$$F_i^j = u^j\big|_i + \delta_i^j \quad (3)$$

where, $\big|_i$ represents covariant derivative with respect to coordinator $x^i$, $\delta_i^j$ is an unit tensor.

Generally, the transformation tensor is non-commutative. Its covariant index and contra-variant index are defined respects with to the initial four-dimensional co-moving coordinator system. Its more strict geometrical theory is under development in some researches (it has some similarity with the concept of geometry algebra, Prof. Chris Doran, Cavendish Laboratory, Cambridge University).

It is clear that the incremental motion of matter is viewed as "deformation" of matter's geometry. So, gravitation field is viewed as matter "deformation" respect with idea vacuum matter. According to Einstein's idear [4-5], the Newton's inertial coordinator system can be replaced by the gradient field of infinitisimal displacement. The equations (2) and (3) show that the displacement gradient determins the current local geometry. Hence, the transformation tensor $F_j^i$ can be named as motion transformation.

Matter motion in four-dimensional spacetime continuum is:

$$\begin{vmatrix} \vec{g}_1 \\ \vec{g}_2 \\ \vec{g}_3 \\ \vec{g}_4 \end{vmatrix} = \begin{vmatrix} 1+u^1\big|_1 & u^2\big|_1 & u^3\big|_1 & u^4\big|_1 \\ u^1\big|_2 & 1+u^2\big|_2 & u^3\big|_2 & u^4\big|_2 \\ u^1\big|_3 & u^2\big|_3 & 1+u^3\big|_3 & u^4\big|_3 \\ u^1\big|_4 & u^2\big|_4 & u^3\big|_4 & 1+u^4\big|_4 \end{vmatrix} \cdot \begin{vmatrix} \vec{g}_1^{\,0} \\ \vec{g}_2^{\,0} \\ \vec{g}_3^{\,0} \\ \vec{g}_4^{\,0} \end{vmatrix} \qquad (4)$$

For moving matter, its current local time gauge is:

$$\vec{g}_4 = u^1\big|_4 \cdot \vec{g}_1^{\,0} + u^2\big|_4 \cdot \vec{g}_2^{\,0} + u^3\big|_4 \cdot \vec{g}_3^{\,0} + (1+u^4\big|_4) \cdot \vec{g}_4^{\,0} \qquad (5)$$

It is clear that local time (attached with matter) measured in standard physical measuring system depends on velocity and time shift rate. Based on definition (1), the local time gauge is:

$$g_{44} = V^2 + (1+u^4\big|_4)^2 \cdot c^2 \qquad (6)$$

For observer, the standard physical measuring system is established in vaccum. If the vaccum is taken as a special matter which depends on cosmic feature where matter is moving, the vaccum can be viewed as refference matter for matter motion. By this understanding, for matter moving in another refference matter field, the initial four-dimensional co-moving coordinator system can be defined by the geometry of refference matter field. In fact, for incremental motion, the refference matter field is the matter itself in initial state.

As the local time gauge in initial matter is:

$$g_{44}^0 = c^2 \qquad (7)$$

The physical time incrementals for matter motion in initial and curent co-moving coordinator systems are, respectively:

$$dt_0 = \sqrt{g_{44}^0} \cdot dx^4 = c \cdot dx^4 \qquad (8)$$

$$dt = \sqrt{g_{44}} \cdot dx^4 = c \cdot \sqrt{(1+u^4\big|_4)^2 + \frac{V^2}{c^2}} \cdot dx^4 \qquad (9)$$

Based on the definition of inertial system, when the $V^i$ is the relative velocity between two inertial system, according to the definition of inertial system, the time gauge must be the same. That is to say one must have:

$$g_{44}^0 = g_{44} \qquad (10)$$

So, the time displacement between initial matter and curent moving matter meets:

$$(1+u^4\big|_4)^2 = 1 - \frac{V^2}{c^2} \qquad (11)$$

When the velocity $V$ is the relative moving speed of inertial coordinator system, as the time gauge is invariant for any inertial systems. To meet the inertial system definition, it means that the current time can be transformed into physical time by factor:

$$\gamma = \frac{1}{\sqrt{1-\frac{V^2}{c^2}}} \qquad (12)$$

This will leads to Lorentz coordinator transformation when it is combined with Newton motion (to be discussed later). This time coordinator transformation factor has been well discussed in special relativity. Here, the paper shows that Lorentz transformation is a natural result of Newton's inertial system concept.

This factor has been widely used to explain the time incremental of moving matter is longer than the time incremental of static matter. However, such an interpretation causes the twin paradox. In this research, the so-called Lorentz time expansion is purely the results of inertial system definition and the principle of the physical laws should be the same in any inertial system.

### 3. Physical Meaning of Matter Motion and Motion Equation

Matter motion, in physical meaning, is measured by an observer with standard physical coordinator system. Such a physical coordinator system is taken as inertial coordinator system in Newton's physics, and physical quantity is well defined in inertial system. For Newton's physics, matter motion is performed in inertial coordinator system defined by idea vacuum. This paper views the idea vacuum as a special matter. So, matter motion will be measured by the deformation of matter with respect to vacuum. Hence, the matter in discussion and the vacuum form a space-time continuum. From this viewpoint to see, the vacuum matter can be replaced by other suitable matter as reference matter.

For Newton's matter-point mechanics, the deformation is defined by the displacement measured in the co-moving coordinator system attached with vacuum matter. For the traditional continuum mechanics, the internal-deformation is defined by the co-moving coordinator system attached with the matter while the absolute deformation motion is described by Newton's matter-point mechanics. For the quantum mechanics, the deformation is defined by the displacement measured in the co-moving coordinator system attached with the quantum existing system.

Based above understanding, matter motion is always defined with reference matter. They form the space-time continuum.

Once the matter motion is well defined by the finite geometric field, the cause of matter motion must be defined. To this purpose, the least action principle raised by Einstein is taken as the basic principle for this research.

As the only field until now is the matter transformation tensor and the action must be a scalar, the (ad hoc) simplest action of general deformation can be defined by the following equation:

$$Action = \int W(u^i|_j) dx^1 dx^2 dx^3 dx^4 \tag{13}$$

where, $W$ is a general function. It can be rewritten as the form:

$$Action = \int W(F^i_j) dx^1 dx^2 dx^3 dx^4 \tag{14}$$

By this way, the broad action of matter is non zero.

The least action principle gives the basic field equation of matter motion as:

$$\left.\frac{\partial W}{\partial (F^i_j)}\right|_j = 0 \tag{15}$$

where, for linear approximation, one has:

$$\frac{\partial W}{\partial (F^i_j)} = \frac{\partial W}{\partial (\widetilde{F}^i_j)} + \frac{\partial^2 W}{\partial (\widetilde{F}^i_j) \partial (\widetilde{F}^k_l)} (F^k_l - \widetilde{F}^k_l) \tag{16}$$

Where, $\tilde{F}_j^i$ is a reference matter motion. The inertial system is an example of reference matter motion. In fact, physically, only incremental motion is measurable in most cases, the reference matter motion is taken as the source for the incremental motion. On this sense, the matter motion equation can be redefined by the following equations:

$$\left.\frac{\partial W}{\partial(\tilde{F}_j^i)}\right|_j = -f^i \qquad (17)$$

where, $f^i$ is named as the action force, produced by environment of matter motion under consideration. By this definition, the incremental motion equation is:

$$\left.\frac{\partial^2 W}{\partial(\tilde{F}_j^i)\partial(\tilde{F}_l^k)}(F_l^k - \tilde{F}_l^k)\right|_j = f^i \qquad (18)$$

Under above definition, the intrinsic physical feature of matter is defined by:

$$C_{jl}^{ik} = \frac{\partial^2 W}{\partial(\tilde{F}_j^i)\partial(\tilde{F}_l^k)} \qquad (19)$$

Usually, one does not care how the original matter motion state is in the absolute spacetime. One just takes the matter under discussion as a self-exist object. In this sense, the initial matter can be generally defined by select the initial co-moving coordinator system to make:

$$\tilde{F}_j^i = \delta_j^i \qquad (20)$$

On this sense, the reference matter is taken to define the spacetime continuum where the matter under discussion is moving.

The principle of physical laws covariant invariance requires the general reference matter motion be coordinator independent. To meet these requirements, the tensor $C_{jl}^{ik}$ must be isotropic tensor. Therefore, one has:

$$C_{jl}^{ik} = \lambda \delta_j^i \delta_l^k + \mu \delta_l^i \delta_j^k \qquad (21)$$

where, $\lambda$ and $\mu$ are the intrinsic feature of matter referring to reference matter. For Newton mechanics, the reference matter is vacuum or ether. For the isotropic tensor in mixture form please refer [6]. In standard physical theory, the mixed form is refused. However, the research shows that the mixture form is necessary to show the non-commutative feature of matter motion transformation tensor. This topic should treat in future rather at present.

Based on the isotropy of inertia matter in spacetime continuum, the inertial matter can be described by two parameters $\lambda_0$ and $\mu_0$ (later the research will show that the inertial mass is defined by parameter $\mu_0$).

For a simple and idea isotropic matter, motion stress tensor $\sigma_j^i$ can be introduced as:

$$\sigma_j^i = (\lambda \delta_j^i \delta_l^k + \mu \delta_l^i \delta_j^k)(F_k^l - \delta_k^l) \qquad (22)$$

Note that it is in mixture tensor form and is different with the form of isotropic tensor in pure covariant or pure anti-covariant form, there three parameters are needed.

If the vacuum is taken as reference matter, then the stress tensor is the stress acting on matter measured by inertial coordinator system (or say, absolute stress). If the initial co-moving coordinator system is a curvature system, the above equations should take their physical component form, as required in tensor theory.

After introducing the motion stress, the matter motion equation (18) becomes:

$$\sigma^i_j\big|_j = f^i \tag{23}$$

where, $f^i$ is the force acting on the space-time continuum from reference matter environment in which the space-time continuum is defined. Geometrically, the traditional convention of positive sign corresponds to the outside normal direction of matter spatial configuration surface (at any time there is a special spatial configuration. This concept is well established by the theory of general relativity. There is no need to introduce the supper-surface concept in higher dimensional spacetime in physics, although it may be very helpful in mathematical sense).

Now, the paper will goes to details. Note that for simplicity, here the initial co-moving coordinator system is taken as the standard physical measuring system defined by equation (1), *so starting from next subsection the conventional partial deferential will be used to replace the covariant differential and physical component of tensor will used when it is appropriate without special mention. The physical component of tensor* $F^i_j$ *is defined by:* $\dfrac{\sqrt{g^0_{(ii)}}}{\sqrt{g^0_{(jj)}}} F^i_j$. *The physical component of covariant derivative is* $\dfrac{\partial}{\sqrt{g^0_{(ii)}}\,\partial x^i}$ . *In the following discussion this rule will be used without notice, as by this way many mathematic operation can be eliminated without damage the main physical consideration.*

### 4. Newton's Equation for Mass-point

For matter motion in vacuum reference, the Newton motion equation can be established. In this paper, inertial matter in traditional physics (Newton mechanics) will be defined as such a matter that the deformation of space-time continuum is symmetrical. The symmetry requirement will guarantees the commutative feature of matter motion. For rigid body without internal spatial deformation, the deformation $F^i_j - \delta^i_j$ is symmetrical. For the pure Newton mass point, the time gradient of time displacement will require the factor defined by equation (12). The condition can be expressed as:

$$\frac{\partial u^1}{c\partial t} = \frac{\partial u^4}{\partial x^1}, \quad \frac{\partial u^2}{c\partial t} = \frac{\partial u^4}{\partial x^2}, \quad \frac{\partial u^3}{c\partial t} = \frac{\partial u^4}{\partial x^3} \tag{24}$$

And the related physical form of deformation tensor and stress tensor are:

$$F^i_j - \delta^i_j = c^{-1} \cdot \begin{vmatrix} 0 & 0 & 0 & V^1 \\ 0 & 0 & 0 & V^2 \\ 0 & 0 & 0 & V^3 \\ V^1 & V^2 & V^3 & 0 \end{vmatrix} \tag{25}$$

$$\sigma^i_j = \mu_0 \cdot c^{-1} \cdot \begin{vmatrix} 0 & 0 & 0 & V^1 \\ 0 & 0 & 0 & V^2 \\ 0 & 0 & 0 & V^3 \\ V^1 & V^2 & V^3 & 0 \end{vmatrix} \qquad (26)$$

Define the inertial mass density as:

$$\rho = \mu_0 \cdot c^{-2} \qquad (27)$$

It interprets the $\mu_0$ as the intrinsic inertial energy of matter.

By equation (23) one gets Newton's mass-point Motion equation:

$$\begin{aligned} f^1 &= \frac{\mu_0}{c} \frac{\partial V^1}{c \partial t} = \rho \frac{\partial V^1}{\partial t} \\ f^2 &= \frac{\mu_0}{c} \frac{\partial V^2}{c \partial t} = \rho \frac{\partial V^2}{\partial t} \\ f^3 &= \frac{\mu_0}{c} \frac{\partial V^3}{c \partial t} = \rho \frac{\partial V^3}{\partial t} \end{aligned} \qquad (28)$$

By the way, the fourth equation:

$$f^4 = \frac{\partial V^1}{\partial x^1} + \frac{\partial V^2}{\partial x^2} + \frac{\partial V^3}{\partial x^3} = 0 \qquad (29)$$

shows that the matter is a rigid body, which is predefined by the deformation without internal spatial deformation. The $f^4 = 0$ is caused by ignoring the time gradient of time displacement.

The above analysis shows that the Newton's matter quality (inertial mass) is determined by the sheer constant $\mu$ between matter and vacuum. Further consideration shows that the sheer constant $\mu$ is Einstein's matter-energy.

In finite geometrical field theory (gauge field theory), the symmetry of motion transformation implies there is no time-space rotation. This rotation-less condition is equivalent with the isotropy of inertial motion. The condition of no rotation between time and space is also the intrinsic feature of Lorentz transformation.

## 5. Newton's Equation in Special Relativity Theory

When the time gauge variation is not ignored, but contribution from the time gradient of time displacement can be ignored, the physical component form of motion deformation of space-time continuum is:

$$F^i_j - \delta^i_j = \gamma \cdot \begin{vmatrix} 0 & 0 & 0 & V^1/c \\ 0 & 0 & 0 & V^2/c \\ 0 & 0 & 0 & V^3/c \\ V^1/c & V^2/c & V^3/c & (u^4|_4) \end{vmatrix} \qquad (30)$$

The stress tensor acting on matter is:

$$\sigma^i_j = \gamma \cdot \mu_0 \cdot c^{-1} \cdot \begin{vmatrix} 0 & 0 & 0 & V^1 \\ 0 & 0 & 0 & V^2 \\ 0 & 0 & 0 & V^3 \\ V^1 & V^2 & V^3 & 0 \end{vmatrix} + (\lambda_0 + \mu_0) \frac{\partial u^4}{\partial t} \delta^i_j \qquad (31)$$

Define the parameter $q$ as:
$$q = (\lambda_0 + \mu_0) \tag{32}$$

By equation (23) one gets Newton's Matter-point Motion equation in special relativity as:

$$f^1 = \gamma \cdot \rho \frac{\partial V^1}{\partial t} + q \frac{\partial^2 u^4}{\partial t \partial x^1}$$
$$f^2 = \gamma \cdot \rho \frac{\partial V^2}{\partial t} + q \frac{\partial^2 u^4}{\partial t \partial x^2} \tag{33}$$
$$f^3 = \gamma \cdot \rho \frac{\partial V^3}{\partial t} + q \frac{\partial^2 u^4}{\partial t \partial x^3}$$

And the fourth equation:

$$f^4 = \gamma \cdot \rho (\frac{\partial V^1}{\partial x^1} + \frac{\partial V^2}{\partial x^2} + \frac{\partial V^3}{\partial x^3}) + q \frac{\partial^2 u^4}{(\partial t)^2} \tag{34}$$

For rigidity of the matter, and the infinitesimal time gradient of time displacement, one still has:

$$f^4 = \gamma \cdot \rho (\frac{\partial V^1}{\partial x^1} + \frac{\partial V^2}{\partial x^2} + \frac{\partial V^3}{\partial x^3}) + q \frac{\partial^2 u^4}{(\partial t)^2} = 0 \tag{35}$$

and motion equations:

$$f^1 = \gamma \cdot \rho \frac{\partial V^1}{\partial t}$$
$$f^2 = \gamma \cdot \rho \frac{\partial V^2}{\partial t} \tag{36}$$
$$f^3 = \gamma \cdot \rho \frac{\partial V^3}{\partial t}$$

This is the form given by special relativity theory.

However, based on this research this is only an approximation. The later research will shows that this approximation cuts out the intrinsic relations among classical mechanics, gravitation field, and quantum mechanics.

## 6. Newton's Gravitation Field and Static Electric Charge Field

For matter in inertial system, when the time and space differentiation of time displacement is commutative, one will has:

$$f^1 = (\gamma \cdot \rho + q) \frac{\partial^2 u^4}{\partial t \partial x^1}$$
$$f^2 = (\gamma \cdot \rho + q) \frac{\partial^2 u^4}{\partial t \partial x^2} \tag{37}$$
$$f^3 = (\gamma \cdot \rho + q) \frac{\partial^2 u^4}{\partial t \partial x^3}$$

If one let:

$$\frac{\partial u^4}{\partial t} = \frac{M}{r}, \quad r = \sqrt{(x^1 - x_0^1)^2 + (x^2 - x_0^2)^2 + (x^3 - x_0^3)^2} \tag{38}$$

Where the parameter $M$ is an constant determined by boundary condition. The fourth equation can automatically produce a zero time force component. That is if:

$$\gamma \cdot \rho [\frac{\partial^2 u^4}{(\partial x^1)^2} + \frac{\partial^2 u^4}{(\partial x^2)} + \frac{\partial^2 u^4}{(\partial x^3)^2}] + q \frac{\partial^2 u^4}{(\partial t)^2} = -4\pi \cdot \gamma \cdot \rho \cdot M \cdot \delta(r) \tag{39}$$

Then, $f^4 = 0$.

It is easy to recognize that the Newton gravity force will require the condition:
$$q = \lambda_0 + \mu_0 = 0 \tag{40}$$
In this case, the one has $M = GM_0$. Here, $G$ is gravitation constant, $M_0$ is the gravity source mass. The boundary condition, in fact, is source condition.

Therefore, for Newton gravitation field, the gravitation mass is identical with the inertial mass, when the constant is determined by Newton gravity field.

For a much broad class of reference matter, the equation (40) can not be true. That will leads to the introduction of electromagnetic field and quantum field, where the meaning of matter is general.

The above analysis shows that the Newton's matter quality is determined by the sheer constant $\mu_0$ between matter and vacuum. The zero volume deformation constant (that is $q=0$) shows that the inertial Newton matter is the simplest matter referring to vacuum.

In general relativity theory, the condition of no rotation between time and space is used to deduce Newton's gravity equation, such as in the Schwarzschild's solution of gravity field [7-8].

It is easy to recognize that the forces:
$$\begin{aligned} f^1 &= q \cdot \frac{\partial^2 u^4}{\partial t \partial x^1} = q \cdot \varepsilon \cdot Q \frac{\partial}{\partial x^1}(\frac{1}{r}) \\ f^2 &= q \cdot \frac{\partial^2 u^4}{\partial t \partial x^2} = q \cdot \varepsilon \cdot Q \cdot \frac{\partial}{\partial x^2}(\frac{1}{r}) \\ f^3 &= q \cdot \frac{\partial^2 u^4}{\partial t \partial x^3} = q \cdot \varepsilon \cdot Q \cdot \frac{\partial}{\partial x^3}(\frac{1}{r}) \end{aligned} \tag{41}$$
define the static electric charge field, if the constant $Q$ is determined by the static electric charge force boundary condition. In classical mechanics, it is defined by introducing test charge. The source parameter $Q$ is the static electric source charge. The parameter $\varepsilon$ is the electric permittivity.

Based on this understanding, the electric charge of matter is defined by:
$$q = \lambda_0 + \mu_0 \tag{42}$$
Hence for Newton matter without the electric charge, one has $\lambda_0 = -\mu_0$. Therefore, the classical Newton matter can be identified by the inertial mass and the static electrical charge.

In a given static electrical field, the Lorentz force for an inertial mass $\rho$ with electric charge $q$ is given by equations:
$$\begin{aligned} f^1 &= \gamma \cdot \rho \frac{\partial V^1}{\partial t} + q\varepsilon Q \frac{\partial}{\partial x^1}(\frac{1}{r}) \\ f^2 &= \gamma \cdot \rho \frac{\partial V^2}{\partial t} + q\varepsilon Q \frac{\partial}{\partial x^2}(\frac{1}{r}) \\ f^3 &= \gamma \cdot \rho \frac{\partial V^3}{\partial t} + q\varepsilon Q \frac{\partial}{\partial x^3}(\frac{1}{r}) \end{aligned} \tag{43}$$

Based on this formulation system, the inertial mass and the static electrical charge are intrinsic features of matter motion in spacetime continuum (in above discussion, the basic reference matter is vacuum or ether). According to Mach's point of view, they are determined by cosmic feature of matter. There is no reason for further explanation. Hence, this research takes them as the ad hoc existence.

Note that the equation (37) can be reduced into equation (36) for macro matter which is defined by: $|q| \ll \rho$, that is: $|\lambda_0 + \mu_0| \ll \dfrac{\mu_0}{c^2}$.

## 7. Motion Equation of Deformable Newton Matter

For simplicity, the spatial deformation of deformable Newton matter is supposed to be principle axes form. The lower index for parameter of spacetime continuum will be omitted. The physical components of the macro formation of deformable matter are:

$$F^i_j - \delta^i_j = \begin{vmatrix} S^1_1 & 0 & 0 & \dfrac{\partial u^4}{\partial x^1} \\ 0 & S^2_2 & 0 & \dfrac{\partial u^4}{\partial x^2} \\ 0 & 0 & S^3_3 & \dfrac{\partial u^4}{\partial x^3} \\ \dfrac{V^1}{c} & \dfrac{V^2}{c} & \dfrac{V^3}{c} & \dfrac{\partial u^4}{\partial t} \end{vmatrix} \tag{44}$$

Where, $S^i_j$ are principle strain components. The corresponding physical stress components are:

$$\sigma^i_j = (\lambda + \mu) \cdot (S^1_1 + S^2_2 + S^3_3 + \dfrac{\partial u^4}{\partial t})\delta^i_j$$

$$+ \mu \cdot \begin{vmatrix} 0 & 0 & 0 & \dfrac{\partial u^4}{\partial x^1} \\ 0 & 0 & 0 & \dfrac{\partial u^4}{\partial x^2} \\ 0 & 0 & 0 & \dfrac{\partial u^4}{\partial x^3} \\ V^1/c & V^2/c & V^3/c & 0 \end{vmatrix} \tag{45}$$

By equation (23), the field matter motion equations are:

$$f^1 = (\lambda + \mu)(\dfrac{\partial^2 u^4}{\partial t \partial x^1} + \dfrac{\partial \Delta}{\partial x^1}) + \dfrac{\mu}{c}\dfrac{\partial^2 u^4}{\partial x^1 \partial t}$$

$$f^2 = (\lambda + \mu)(\dfrac{\partial^2 u^4}{\partial t \partial x^2} + \dfrac{\partial \Delta}{\partial x^2}) + \dfrac{\mu}{c}\dfrac{\partial^2 u^4}{\partial x^2 \partial t}$$

$$f^3 = (\lambda + \mu)(\dfrac{\partial^2 u^4}{\partial t \partial x^3} + \dfrac{\partial \Delta}{\partial x^3}) + \dfrac{\mu}{c}\dfrac{\partial^2 u^4}{\partial x^3 \partial t} \tag{46}$$

$$f^4 = \dfrac{\lambda + \mu}{c}[\dfrac{\partial^2 u^4}{(\partial t)^2} + \dfrac{\partial \Delta}{\partial t}] + \dfrac{\mu}{c}[\dfrac{\partial V^1}{\partial x^1} + \dfrac{\partial V^2}{\partial x^2} + \dfrac{\partial V^3}{\partial x^3}]$$

where,

$$\Delta = (S^1_1 + S^2_2 + S^3_3) \tag{47}$$

is the volume variation of infinitesimal spatial deformation of matter. It shows that macro spatial deformation of matter field is related with quantum mechanics.

Now, we examines its macro-behavior for Newton's matter field. Here, we suppose that the derivative of time displacement is commutative.

For Newton's matter motion, one has:

$$\dfrac{\partial u^4}{\partial x^i} = \dfrac{V^i}{c} \tag{48}$$

The cosmic force component is:

$$f^4 = \frac{\lambda + \mu}{c}[\frac{\partial^2 u^4}{(\partial t)^2} + \frac{\partial \Delta}{\partial t}] + \mu[\frac{\partial^2 u^4}{(\partial x^1)^2} + \frac{\partial^2 u^4}{(\partial x^2)^2} + \frac{\partial^2 u^4}{(\partial x^3)^2}] \quad (49)$$

If the cosmic force component is zero, that is if the cosmic is idea vacuum, one will get the motion equation for time displacement field:

$$-\frac{\lambda + \mu}{c}[\frac{\partial^2 u^4}{(\partial t)^2} + \frac{\partial \Delta}{\partial t}] = \mu[\frac{\partial^2 u^4}{(\partial x^1)^2} + \frac{\partial^2 u^4}{(\partial x^2)^2} + \frac{\partial^2 u^4}{(\partial x^3)^2}] \quad (50)$$

There are four typical matter motion will be discussed bellow. This in fact is a free matter model.

## *7. 1 Non-compressible Newton's Matter Motion Equation*

The following research will show that the non-compressible Newton's matter behaves like wave. So, the matter in its intrinsic meaning has wave-particle duality. The non-compressible Newton matter is defined as:

$$\Delta = 0 \quad (51)$$

For non-compressible Newton's matter, the time displacement meets equation:

$$\frac{\partial^2 u^4}{(\partial x^1)^2} + \frac{\partial^2 u^4}{(\partial x^2)^2} + \frac{\partial^2 u^4}{(\partial x^3)^2} = -\frac{\mu}{\lambda + \mu}\frac{\partial^2 u^4}{c(\partial t)^2} \quad (52)$$

This is wave equation of the quantum matter with inertial mass in reference matter space-time continuum.

**(a).** For the case: $\lambda + \mu < 0$, $\mu > 0$ its wave solution of equation (52) is:

$$u^4 = U_0 \exp[(\vec{R} \cdot \vec{k} \pm \omega t)\tilde{j}] \quad (53)$$

where, $U_0$ is a constant, to be determined by boundary condition; $\tilde{j}$ is the sign of imaginary number; $\vec{k}$ is wave number spatial vector; $\vec{R}$ is spatial position vector; $\omega$ is the frequency of deformable matter. There is a wave number-frequency relation equation:

$$k^2 = \frac{\mu}{-(\lambda + \mu)c}\omega^2 = \frac{\omega^2}{v_-^2} \quad (54)$$

Here, the wave speed is defined as:

$$v_- = \sqrt{-c \cdot \frac{\lambda + \mu}{\mu}} \quad (55)$$

The matter spatial forces are:

$$f^i = (\lambda + \mu + \frac{\mu}{c})\frac{\partial^2 u^4}{\partial t \partial x^i} = -\tilde{\rho} U_0 \omega k_i \exp[(\vec{R} \cdot \vec{k} \pm \omega t)\tilde{j}] \quad (56)$$

For vacuum reference matter,

$$\tilde{\rho} = (\lambda + \mu + \frac{\mu}{c}) \quad (56')$$

$\tilde{\rho}$ is the equivalent mass in Einstein's mass-energy equation. Note that when the electrical charge is near zero, the equivalent mass is near Newtonian mass.

**(b).** For the case: $\lambda + \mu > 0$, $\mu > 0$, the wave-particle solution of equation (52) is:

$$u^4 = U_0 \exp(\pm \vec{R} \cdot \vec{l} \pm \omega t \tilde{j}) \quad (57)$$

where, $U_0$ is a constant, to be determined by boundary condition; $\tilde{j}$ is the sign of imaginary number; $\vec{l}$ is spatial motion direction vector; $\vec{R}$ is spatial position vector. There is a particle size-frequency relation equation:

$$l^2 = \frac{\mu}{(\lambda+\mu)c}\omega^2 = \frac{\omega^2}{v_+^2} \tag{58}$$

Here, the $l$ represents typical size of the matter. The parameter $v_+$ relates the typical size and the frequency, it is easy to see that $v_+ = \sqrt{c\frac{\lambda+\mu}{\mu}}$. Its full discussion will be given in quantum mechanics. The sign before the spatial motion vector is to make the solution be limit.

The matter spatial forces are:

$$f^i = \tilde{\rho}\frac{\partial^2 u^4}{\partial t \partial x^i} = \pm \tilde{j}\tilde{\rho}U_0 \omega l_i \exp(\pm\vec{R}\cdot\vec{l} \pm \omega t j) \tag{59}$$

The Newton's force is localized. Hence, Newton's matter behaves as vibrating-particle in quantum mechanics. That is the wave-particle duality.

(c). For the case: $\lambda + \mu = 0$, $\mu > 0$ the conservative field solution of equation (52) is:

$$u^4 = \frac{U_0}{r} \tag{60}$$

The matter spatial force is gravity force or electric charge force. It returns to the cases discussed for simple matter in the earlier sections.

Summering up above discussion, it can be concluded that the wave-particle solution and the wave solution in quantum mechanics are also determined by the two basic parameters $(\lambda, \mu)$. Therefore, unified field theory can be achieved by the basic parameters of matter under the principle of spacetime continuum isotropy and the least action principle.

## 7.2 Compressible Newton's Matter Motion Equation

The following research will show that the macro-deformation of Newton's matter has intrinsic stress-strain relation equations.

For compressible Newton's matter, in the case that the cosmic force is zero, that is $f^4 = 0$, the fourth equation of the equations (46) gives:

$$\frac{\lambda+\mu}{c}[\frac{\partial^2 u^4}{(\partial t)^2} + \frac{\partial \Delta}{\partial t}] + \frac{\mu}{c}[\frac{\partial V^1}{\partial x^1} + \frac{\partial V^2}{\partial x^2} + \frac{\partial V^3}{\partial x^3}] = 0 \tag{61}$$

It shows that the macro-spatial deformation of matter will produce time displacement. One gets:

$$\frac{\partial^2 u^4}{(\partial t)^2} = -\frac{\partial \Delta}{\partial t} - \frac{\mu}{\lambda+\mu}[\frac{\partial V^1}{\partial x^1} + \frac{\partial V^2}{\partial x^2} + \frac{\partial V^3}{\partial x^3}] \tag{62}$$

For Newton's matter, the basic solution is:

$$\frac{\partial u^4}{\partial t} = -\frac{\lambda+2\mu}{\lambda+\mu}[\frac{\partial u^1}{\partial x^1} + \frac{\partial u^2}{\partial x^2} + \frac{\partial u^3}{\partial x^3}] = -\frac{\lambda+2\mu}{\lambda+\mu}\Delta \tag{63}$$

In this case, the macro-spatial deformation of Newtonian matter will cause time displacement gradient. This will produce quantum motion. Therefore, the macro deformation is determined by the intrinsic structure of matter composition. This is because that from the definition of Newtonian matter, one gets:

$$\frac{\partial^2 u^4}{\partial t \partial x^i} = -\frac{\lambda + 2\mu}{\lambda + \mu} \frac{\partial \Delta}{\partial x^i}, \quad i = 1,2,3 \tag{64}$$

Hence, one gets the spatial forces are:

$$f^i = (\lambda + \mu)(\frac{\partial^2 u^4}{\partial t \partial x^i} + \frac{\partial \Delta}{\partial x^2}) + \frac{\mu}{c} \frac{\partial^2 u^4}{\partial x^i \partial t} \tag{46}$$

That is, when the cosmic force component is zero, the spatial force is:

$$f^i = -(\mu + \frac{\mu}{c} \cdot \frac{\lambda + 2\mu}{\lambda + \mu}) \frac{\partial \Delta}{\partial x^i}, \quad i = 1,2,3 \tag{65}$$

It gives the relation for matter macro spatial volume deformation in space-time continuum with spatial force from cosmic background matter.

In engineering, the macro-matter itself is taken as the reference matter in continuum mechanics. The physical meaning of above discussion is to show that to describe macro continuum deformation, the two basic parameters can be obtained by the microphysics ways. In essential sense, it shows that the macro deformation parameters ($\tilde{\lambda}, \tilde{\mu}$) can be derived by basic parameters and micro matter components motion. For more complicated motion, that is the cosmic force is not zero, will not be performed in this paper.

**8. Zero Inertial Mass Matter Motion Equation**

Zero Newton's mass field matter can be defined by $\mu = 0$. For zero mass matter fields, such as light quantum or electromagnetic field, the equation can be simplified. It will show that the volume deformation is possible. As: $\mu = 0$. The equations (46) become as:

$$\begin{aligned} f^1 &= \lambda(\frac{\partial^2 u^4}{\partial t \partial x^1} + \frac{\partial \Delta}{\partial x^1}) \\ f^2 &= \lambda(\frac{\partial^2 u^4}{\partial t \partial x^2} + \frac{\partial \Delta}{\partial x^2}) \\ f^3 &= \lambda(\frac{\partial^2 u^4}{\partial t \partial x^3} + \frac{\partial \Delta}{\partial x^3}) \\ f^4 &= \frac{\lambda}{c}[\frac{\partial^2 u^4}{(\partial t)^2} + \frac{\partial \Delta}{\partial t}] \end{aligned} \tag{66}$$

For free matter field, that is the spatial forces are all zeros, one gets:

$$\frac{\partial^2 u^4}{\partial x^i \partial t} = -\frac{\partial \Delta}{\partial x^i}, \quad i = 1,2,3 \tag{67}$$

As the equation is derived on the condition of force-less and mass-less, so it is the intrinsic geometrical feature of the spacetime continuum. In fact, this solution is implied by the Schwarzschild's solution of Einstein's gravity field theory. Hence, Einstein has enough reason to view the gravity field as the curvature of space-time continuum.

If the matter behaves as Newton's matter, it becomes:

$$\frac{\partial V^i}{c \partial t} = -\frac{\partial \Delta}{\partial x^i}, \quad i = 1,2,3 \tag{68}$$

For light, as its speed is invariant, the equation shows that the light will have no spatial variation. For electromagnetic field in medium, the equation shows that the permittivity and the magnetic permeability (their conduct determines the velocity) are related with the curvature of space-time continuum

For cosmic force free mass-less field, the fourth equation of (66) gives equation:

$$\frac{\partial u^4}{\partial t} = -\frac{\partial u^1}{\partial x^1} - \frac{\partial u^2}{\partial x^2} - \frac{\partial u^3}{\partial x^3} \tag{69}$$

This equation shows that if the spacetime is defined by mass-less matter motion, the space and time must have intrinsic relation. This is the underlying reason for the correctness of Lorentz transformation for electromagnetic field.

This can be simply explained as the incompressibility of spacetime as a cosmic matter. This in essential sense is the base of Lorentz transformation in vacuum. However, based on this research, the Lorentz transformation can not be taken as the bases for general matter fields.

## 9. Born Rigidity and Lorentz Transformation

The definition of inertial system will require the Born rigidity. That is to say the distances of material particles remain constant with respect to the instantaneously co-moving inertial system when the material is subjected to a constant proper acceleration. Such a rigidity requirement should be physically sound in average sense. In the finite geometry field, according to equation (4), for any time duration $T_{ab}$ between two temporal points A and B, the Born rigidity can be expressed as:

$$\frac{1}{T_{ab}}\int_{a^4}^{b^4}\sqrt{g_{11}}\,dx^4 = \frac{1}{T_{ab}}\int_{a^4}^{b^4}\sqrt{g_{11}^0 + (u^4|_1)^2 c^2}\,dx^4 = 1$$

$$\frac{1}{T_{ab}}\int_{a^4}^{b^4}\sqrt{g_{22}}\,dx^4 = \frac{1}{T_{ab}}\int_{a^4}^{b^4}\sqrt{g_{22}^0 + (u^4|_2)^2 c^2}\,dx^4 = 1 \tag{70}$$

$$\frac{1}{T_{ab}}\int_{a^4}^{b^4}\sqrt{g_{33}}\,dx^4 = \frac{1}{T_{ab}}\int_{a^4}^{b^4}\sqrt{g_{33}^0 + (u^4|_3)^2 c^2}\,dx^4 = 1$$

For standard initial coordinator system $g_{ij}^0 = \delta_{ij}$, that is taken $c=1$. Based on the isotropy requirement of motion transformation for inertial motion, the related condition equations are:

$$u^4|_1 = u^1|_4, \quad u^4|_2 = u^2|_4, \quad u^4|_3 = u^3|_4 \tag{71}$$

On these conditions, the space displacement fields are given by equations:

$$ds_{11} = \int_{a^4}^{b^4} u^1 \sqrt{1+(u^4|_1)^2}\,dx^4 = \int_{a^4}^{b^4} u^1 \sqrt{1+(u^1|_4)^2}\,dx^4$$

$$ds_{22} = \int_{a^4}^{b^4} u^2 \sqrt{1+(u^4|_2)^2}\,dx^4 = \int_{a^4}^{b^4} u^2 \sqrt{1+(u^2|_4)^2}\,dx^4 \tag{72}$$

$$ds_{33} = \int_{a^4}^{b^4} u^3 \sqrt{1+(u^4|_3)^2}\,dx^4 = \int_{a^4}^{b^4} u^3 \sqrt{1+(u^3|_4)^2}\,dx^4$$

Their spatial shortest-distance path solutions are:

$$u^1 = C^1 \cdot \cosh(\frac{x^4 - C_0}{C^1})$$

$$u^2 = C^2 \cdot \cosh(\frac{x^4 - C_0}{C^2}) \tag{73}$$

$$u^3 = C^3 \cdot \cosh(\frac{x^4 - C_0}{C^3})$$

where $C^1, C^2, C^3$ are the displacement at time $C_0$. For infinitesimal time coordinator interval,

one will get the approximation:

$$\frac{1}{3}\{[(u^1)^2 - (C^1)^2] + [(u^2)^2 - (C^2)^2] + [(u^3)^2 - (C^3)^2]\} = (x^4 - C_0)^2 \tag{74}$$

It means that the square of average incremental displacement equals to the square of time incremental interval. In form, it can be rewritten as the local geometrical invariance form in coordinator notation as:

$$(dx^1)^2 + (dx^2)^2 + (dx^3)^2 = (dx^4)^2 \tag{75}$$

Therefore, the Lorentz transformation is derived from the definition of inertial system for rigid material. That is to say, the Born rigidity of inertial system definition will produce the Lorentz transformation.

It is easy to find the velocity components are:

$$V^1 = u^1\big|_4 = \sinh(\frac{x^4 - C_0}{C^1})$$

$$V^2 = u^2\big|_4 = \sinh(\frac{x^4 - C_0}{C^2}) \tag{76}$$

$$V^3 = u^3\big|_4 = \sinh(\frac{x^4 - C_0}{C^3})$$

Physically, the upper speed is limited by the light speed, hence one has:

$$-1 < \sinh(\frac{x^4 - C_0}{C^i}) < 1 \tag{77}$$

This can be approximated as:

$$-1 < (\frac{x^4 - C_0}{C^i}) < 1 \tag{78}$$

The acceleration components are:

$$A^1 = \frac{1}{C^1}\cosh(\frac{x^4 - C_0}{C^1})$$

$$A^2 = \frac{1}{C^2}\cosh(\frac{x^4 - C_0}{C^2}) \tag{79}$$

$$A^3 = \frac{1}{C^3}\cosh(\frac{x^4 - C_0}{C^3})$$

Therefore, to meet the Born rigidity the material must be accelerated by special ways. They are time-dependent, where the spatial displacements are parameters.

The upper limit of light speed makes the acceleration can be approximated as:

$$A^i = \frac{1}{C^i}[1 + \frac{1}{2}(\frac{x^4 - C_0}{C^i})^2] \tag{80}$$

By the definition of energy done by the force, one has:

$$E = \int \sum_{i=1,2,3}[A^i u^i]dx^4 = \int \sum_{i=1,2,3}[\cosh(\frac{x^4 - C_0}{C^i})]^2 dx^4 \tag{81}$$

To make the energy be limit value, the parameter $C^1, C^2, C^3$ must be pure imaginary number. In this case, the displacement given by the spatial short-distance path solutions are complex number. In this case, it is called wave function in quantum mechanics. The above equation is taken as the normality condition.

Based on above analysis, it can be concluded that the quantum mechanics can be traced back to the same principle as the classical mechanics. Combining with the conclusion about special relativity, it is clear the definition of inertial system is essential for the whole physics. The grand unified theory should base on the concept of inertia and the principle of general relativity.

**10. Conclusion**

The research shows an unified field theory of physics can be established with the idea introducing motion transformation, which is completely determined by the displacement gradient in four-dimensional space-time continuum. The vacuum is taken as a special matter with one intrinsic physical parameter-light speed $c$. For any matter, it can be identified with co-moving coordinators with local varying geometry. The local varying geometry represents the motion of matter in consideration. The motion equation of matter is determined by cosmic environment where the matter existing, where two intrinsic physical parameters are introduced to define matter as $(\lambda, \mu)$.

When the time gradient of time displacement is forced to be zero, the Newton's physics for matter is gotten. It shows that the Newton's mass is determined by the shear feature of space-time continuum. The Newton's mass is expressed as $\rho = \mu/c^2$.

When the matter has no macro spatial motion, such as gravity field or electronic field, the time displacement field is introduced to describe the matter motion. One finds that gravity field, electro-magnetic field, and quantum field are time gradient field. They are related with Newton's mechanics in intrinsic sense. The absolute electrical charge quantity may be expressed as $q = (\lambda + \mu)$. When there is no electrical charge quantity, the gravity mass and the inertial mass in Newton's mechanics are the same.

The matter has three typical existing forms: traveling waves, localized harmonic vibrating particle, and exponential expanding or decaying matter field.

The commutable matter motion defines conservative field. The non-commutable matter motion defines quantum field, where $\tilde{\rho} = (\lambda + \mu + \frac{\mu}{c})$ is defined as equivalent mass in Einstein's mass-energy equation. The wave particle duality is explained by deformable matter motion. The observation of cosmic expanding and the gravity field matter in three-dimensional space can be explained. Finely, the paper introduces the Born rigidity of inertial motion to derive the Lorentz transformation.

It can be concluded that the quantum mechanics can be traced back to the same principle as the classical mechanics. Combining with the conclusion about special relativity, it is clear the definition of inertial system is essential for the whole physics. The grand unified theory should base on the concept of inertia and the principle of general relativity. So, an unified physics of matter motion can be expressed by the finite geometrical field theory.

**Refference**